\documentclass[preprint,showpacs,preprintnumbers,amsmath,amssymb,superscriptaddress,aps,prb]{revtex4}

\usepackage{graphicx}
\usepackage{dcolumn}
\usepackage{bm}
\usepackage{longtable}
\usepackage{multirow}
\usepackage{amssymb}
\usepackage{dcolumn}

\begin{document}

\preprint{}

\title{Interface energies of $(100)_{\rm YSZ}$ and $(111)_{\rm YSZ}$ epitaxial
islands on $(0001)_{\alpha-{\rm{}Al}_2{\rm{}O}_3}$ substrates from first principles}

\author{F. Lallet}
\email{f_lallet@ensci.fr}
\affiliation{Laboratoire de Sciences des Proc\'ed\'es C\'eramiques et Traitements
de Surface, UMR-CNRS 6638, Ecole Nationale Sup\'erieure de C\'eramiques
Industrielles, 47 avenue Albert Thomas, 87065 Limoges cedex, France}
\author{N. Olivi-Tran}
\email{n_olivi-tran@ensci.fr}
\affiliation{Laboratoire de Sciences des Proc\'ed\'es C\'eramiques et Traitements
de Surface, UMR-CNRS 6638, Ecole Nationale Sup\'erieure de C\'eramiques
Industrielles, 47 avenue Albert Thomas, 87065 Limoges cedex, France}
\author{\firstname{Laurent J.} \surname{Lewis}}
\email{Laurent.Lewis@UMontreal.CA}
\affiliation{D\'{e}partement de Physique et Regroupement Qu\'{e}b\'{e}cois
sur les Mat\'{e}riaux de Pointe (RQMP), Universit\'{e} de Montr\'{e}al, C.P.
6128, Succursale Centre-Ville,\\ Montr\'{e}al, Qu\'{e}bec, Canada H3C 3J7}

\date{\today}

\begin{abstract}

We present an {\em ab initio} study of the interface energies of cubic
yttria-stabilized zirconia (YSZ) epitaxial layers on a
$(0001)_{\alpha-{\rm{}Al}_2{\rm{}O}_3}$ substrate. The interfaces are
modelled using a supercell geometry and the calculations are carried out in
the framework of density-functional theory (DFT) and the local-density
approximation (LDA) using the projector-augmented-wave (PAW) pseudopotential
approach. Our calculations clearly demonstrate that the $(111)_{\rm YSZ} ||
(0001)_{\alpha-{\rm{}Al}_2{\rm{}O}_3}$ interface energy is lower than that of
$(100)_{\rm YSZ} || (0001)_{\alpha-{\rm{}Al}_2{\rm{}O}_3}$. This result is
central to understanding the behaviour of YSZ thin solid film islanding on
$(0001)_{\alpha-{\rm{}Al}_2{\rm{}O}_3}$ substrates, either flat or in
presence of defects.

\end{abstract}

\vfill
\maketitle


\section{Introduction}
\label{intro}

Over the last few years, many experimental efforts have been expended on the
fabrication of self-patterned, epitaxial nanocrystals (metallic,
semiconductor or oxyde) on crystalline substrates.\cite{Repain, Silly,
Alchalabi, Nie, Sanchez, Vasco} The aim is to synthesize homogeneous patterns
of epitaxial crystals in order to induce quantum confinement --- intimately
related to the shape and the size of the nanocrystals --- in order to achieve
enhanced optical and/or magnetic properties.\cite{He, Heiss, Zhao, Prokop,
Tasco, Klimov, Wenisch} Several theoretical investigations have been
concerned with the physical parameters responsible for the geometric
properties of the nanocrystals.\cite{Nurminen, Kalke, Liu1, Kawamura, Russo,
Lallet2}

The fabrication of such systems can be realised through various techniques
involving the formation of nanometer-scale islands in a collective way, a
process known as self-organisation. The basic idea is to promote the
formation of nanocrystals on a crystalline substrate using thin solid films
which demonstrate spontaneous evolution from a continuous 2D solid layer to a
rough and/or discontinuous film, i.e., 3D epitaxial nanocrystals. The
formation of the nanocrystals takes place during or after the deposition of
the film as a way of reducing the total energy of the epitaxial
$\{{\rm{}layer}||{\rm{}substrate}\}$ system by the relaxation of the
interface and/or surface stresses and strains. We briefly describe, in what
follows, three of the most popular ``bottom-up'' approaches involved in
epitaxial nanocrystals self-organisation processes.

A first approach is chemical vapor deposition (CVD), which consists in mixing
chemical species in vacuum, which then react or decompose on the surface of a
substrate to form a thin solid film. A second approach is physical vapor
deposition (PVD), whereby matter is extracted from a solid target with, for
example, a laser or an ion beam; the extracted ions attach to the surface of
a substrate and eventually constitute a thin solid film. CVD and PVD, which
lead to the formation of nanocrystals during the deposition process, have
been successfully applied to the fabrication of self-organized arrays of
semiconductor or metal nanocrystals, commonly called quantum dots (QDs). One
of the most widely studied system is $\{{\rm{}Si}||{\rm{}Ge}\}$; this is
characterized by the formation of epitaxial, faceted Ge QDs either at the top
of a continuous Ge wetting layer (Stranski-Krastanov growth), or directly at
the surface of the Si substrate (Volmer-Weber growth).\cite{Sutter, Tromp,
Capellini, Portavoce, Chen, Wagner1, Wagner2, Voigtlander} A third approach
goes by the deposition of a continuous thin xerogel film at the surface of a
substrate by sol-gel dip-coating.\cite{Brinker} Through thermal treatment,
the continuous thin solid film crystallizes and breaks into several crystals
through surface diffusion,\cite{Olivi1, Olivi2, Lallet1} leading to the
formation of discrete epitaxial islands on the surface of the substrate. In
this case, unlike PVD or CVD, the formation of the epitaxial islands takes
place after the deposition of the film. This technology is particularly
efficient for designing arrays of oxide nano-islands, as recently
demonstrated by Bachelet.\cite{Bachelet}

In this article, we are concerned with the $\{$${\rm YSZ} ||
(0001)_{\alpha-{\rm{}Al}_2{\rm{}O}_3}$$\}$ system, where epitaxial
nano-islands of cubic yttria-stabilized zirconia (YSZ) form during the
thermal treatment of a thin xerogel film deposited by sol-gel dip-coating on
a $(0001)_{\alpha-{\rm{}Al}_2{\rm{}O}_3}$ substrate. Experimentally, Bachelet
{\em et al}.\ have demonstrated that the shape and size of the YSZ islands
are directly linked to their epitaxial relation with the
substrate.\cite{Bachelet} Indeed, for a substrate without defects, the
islands are top-flat with large interface areas and exhibit the following
in-plane and out-of-plane crystallographic orientations:
   \begin{eqnarray}
   (100)_{\rm YSZ} || (0001)_{\alpha-{\rm{}Al}_2{\rm{}O}_3}, & & [001]_{\rm YSZ} || [010]_{\alpha-{\rm{}Al}_2{\rm{}O}_3}, \\
   (100)_{\rm YSZ} || (0001)_{\alpha-{\rm{}Al}_2{\rm{}O}_3}, & & [001]_{\rm YSZ} || [110]_{\alpha-{\rm{}Al}_2{\rm{}O}_3},
   \end{eqnarray}
whereas for a substrate containing defects, some islands are round and thicker
than the top-flat ones but with lower interface areas,
and possess the in-plane and-out-of plane orientations:
   \begin{equation}
   (111)_{\rm YSZ} || (0001)_{\alpha-{\rm{}Al}_2{\rm{}O}_3}, \;\; [1 \bar{1}0]_{\rm YSZ} || [110]_{\alpha-{\rm{}Al}_2{\rm{}O}_3}.
   \end{equation}
The morphological evolution of the thin solid film into discrete nano-islands
proceeds by an abnormal grain growth driven by the interface during thermal
treatment.\cite{Lange} From our previous theoretical investigations of this
system,\cite{Lallet1} and in good agreement with experimental results, we
demonstrated that the shape and size transition from top-flat to round is
linked to the presence of defects at the surface of the substrate which
induce enhanced growth in height.

However, the preferred formation of interfaces (1) and (2) over interface (3)
for a perfect $(0001)_{\alpha-{\rm{}Al}_2{\rm{}O}_3}$ substrate is not
clearly understood. On the basis of energy considerations it can be argued
that, to first order, an epitaxial crystal is in equilibrium with both the
vacuum through the free surface energy and with the substrate through the
interface energy, where both energies are related to the crystallographic
orientations. Using {\em ab initio} methods, Ballabio {\em et al}.\ have
demonstrated that the free surface energy of $(100)_{\rm YSZ}$ is higher than
that of $(111)_{\rm YSZ}$.\cite{Ballabio}

In this article, we propose to examine the interface energies defined by the
epitaxial relations (1), (2), and (3) on a perfect
$(0001)_{\alpha-{\rm{}Al}_2{\rm{}O}_3}$ substrate. We argue that the
knowledge of the interface energies is sufficient for a proper comparison of
the behaviour of the three interfaces. There are several theoretical
investigations of $\{{\rm{}metal}||{\rm{}oxide}\}$ interfaces, in particular
for $(0001)_{\alpha-{\rm{}Al}_2{\rm{}O}_3}$ because of its technological
significance in thermal barrier coatings and catalytic devices.\cite{Duffy,
Zhang1, Batyrev1, Zhang2, Siegel, Sinnott, Dmitriev, Christensen3} However,
few studies have been devoted to $\{{\rm{}oxide}||{\rm{}oxide}\}$ interfaces
\cite{Christensen2} and, to the best of our knowledge, none have been
concerned with the $\{$${\rm YSZ} ||
(0001)_{\alpha-{\rm{}Al}_2{\rm{}O}_3}$$\}$ system. Our calculations clearly
demonstrate that interface (3) is energetically favored over interfaces (1)
and (2). We therefore propose a general explanation for the behaviour of the
islanding process of YSZ thin solid films on
$(0001)_{\alpha-{\rm{}Al}_2{\rm{}O}_3}$ substrates, either perfect or with
surface defects, in the light of experimental and theoretical
investigations.\cite{Bachelet, Lallet1}

The article is constructed as follows. In section \ref{numerical} we give the
details of the numerical procedure, followed in section \ref{structure} by a
demonstration of the ability of the PAW pseudopotentials to reproduce the
correct structural properties of $\alpha$ and $\kappa$-Al$_2$O$_3$, the
low-pressure polymorphs of ZrO$_2$, and the Y$_2$O$_3$ bixbyite structure. In
section \ref{unrelaxed} we first discuss the calculation of the unrelaxed and
relaxed stoichiometric free surface energies, then we present the
atomic-scale models for the the (1), (2) and (3) $\{$${\rm YSZ} ||
(0001)_{\alpha-{\rm{}Al}_2{\rm{}O}_3}$$\}$ interfaces and the results of our
calculations. A general conclusion is provided in Section \ref{conclusion}.


\section{Computational details}
\label{numerical}

All calculations were carried out in the framework of density functional
theory (DFT) using the {\em Abinit} code,\cite{Abinit} where the wave
functions are expanded in plane waves. The atomic pseudopotentials were
constructed with the {\em atompaw} program\cite{atompaw} within the
frozen-core approximation, using the projector-augmented-waves (PAW) method
originally proposed by Bl\"ochl.\cite{Blochl} For the exchange-correlation
functional, we employed the local density approximation (LDA) as parametrized
by Perdew and Wang.\cite{Perdew1} The atomic wave functions were augmented
with 3, 6, 5, and 5 projectors within a spherical augmentation region of
radii 1.4, 1.8, 2, and 2 Bohrs for O, Al, Y, and Zr atoms, respectively. The
$2s$ and $3s$ semi-core states of Al, as well as the $4s$ and $4p$ semi-core
states of Y and Zr, were treated as valence states to generate the
pseudo-wave and projector functions within the augmentation region. We found
that taking Y and Zr semi-core states as valence is debatable. Indeed, Jansen
\cite{Jansen} demonstrated that, due to the large energy difference between O
and Zr$(4s,4p)$ resonances, the O-Zr$(4s,4p)$ hybridization is weak in
ZrO$_2$. This argument was applied by Christensen and Carter to study the
free surfaces of ZrO$_2$ low-pressure polymorphs\cite{Christensen1} and the
$\{$$(001)_{ZrO_2} || (10\bar{1}2)_{\alpha-{\rm{}Al}_2{\rm{}O}_3}$$\}$
interface.\cite{Christensen2} However, in previous studies of bulk YSZ by
Stapper {\em et al}.\cite{Stapper} and of YSZ slabs by Ballabio et
al.,\cite{Ballabio} the $(4s,4p)$ semi-core states of Y and Zr were treated
as valence. Here, we are also dealing with YSZ bulk and slab structures, as
it is our purpose to characterize the $\{$${\rm YSZ} ||
(0001)_{\alpha-{\rm{}Al}_2{\rm{}O}_3}$$\}$ interfaces. In section
\ref{structure} we will demonstrate that the treatment of $4s$ and $4p$
semi-core states of Y and Zr as valence states is appropriate to accurately
describe the structural properties of Y$_2$O$_3$ bixbyite and ZrO$_2$
low-pressure polymorphs.

Additional details are as follows. For the $\alpha$ and $\kappa$ phases of
Al$_2$O$_3$, as well as the ZrO$_2$ low-pressure polymorphs and the
Y$_2$O$_3$ bixbyite structure, all discussed in section \ref{structure}, we
used a kinetic energy cutoff of 15 Ha ($\approx$ 408 eV) and
$2\times2\times2$ Monkhorst-Pack grid\cite{Monkhorst} for the Brillouin-zone
integrations of bulk unit cells, which is a standard choice for wide band gap
oxides.\cite{Christensen2} With these parameters, the total energies are
converged to within $10^{-2}$ Ha/atom (0.2 eV/atom) and the forces to better
than $10^{-4}$ Ha/(Bohr.atom). The atoms were relaxed to their ground-state
positions using the Broyden-Fletcher-Goldfarb-Shanno (BFGS)
algorithm.\cite{Broyden}

The slab geometry for the $(0001)_{\alpha-{\rm{}Al}_2{\rm{}O}_3}$ system is
described in section \ref{unrelaxed}; in this case we used a
$2\times2\times1$ Monkhorst-Pack grid in the $z$ direction. For the (1), (2)
and (3) $\{$${\rm YSZ} || (0001)_{\alpha-{\rm{}Al}_2{\rm{}O}_3}$$\}$
interfaces, and according to previous studies,\cite{Ballabio, Stapper} only
the $\Gamma$ point was used to integrate the Brillouin zone. The convergence
criteria for relaxation were the same as above.


\section{Structural parameters of the bulk phases}
\label{structure}

We present here our results for the various bulk phases in order to ensure
that our approach yields the correct structural parameters. For this purpose,
we computed the relaxed lattice parameters and ionic positions (we follow the
Wyckoff convention\cite{Wyckoff}) for each crystalline structure. We also
computed the ground state energies $E$ as a function of the volume $V$ of the
unit cell; these can be fitted to the Murnaghan equation of
state:\cite{Murnaghan}
   \begin{equation}
   E(V)=E_0+\left( \frac{B_0 V}{B^{'}_0} \right) \left( \frac{(V_0/V)^{B^{'}_0}}{B^{'}_0 -1} + 1 \right) -
             \frac{B_0 V_0}{B^{'}_0-1}
   \end{equation}
with
   \begin{equation}
   B^{'}_0 = \left( \frac{dB_0}{dP} \right)_{(P=0)},
   \end{equation}
thus yielding the equilibrium volume $V_0$ and the bulk modulus $B_0$;
$P$ is the pressure.

\begin{figure}[tb]
\includegraphics[width=8cm]{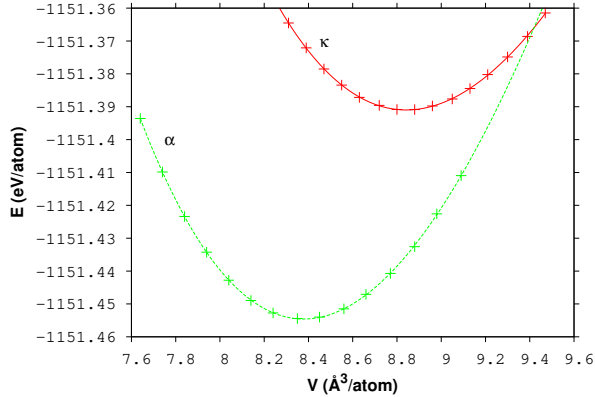}
\caption{(Color online) Energy vs volume for $\alpha$-Al$_2$O$_3$ and
$\kappa$-Al$_2$O$_3$, as indicated}
\label{Fig_E_vs_V_al2o3}
\end{figure}

$\alpha$-Al$_2$O$_3$ (also known as corundum) is the thermodynamically stable
polymorph of alumina at low pressure. It is widely used for epitaxial thin
solid film growth, in particular the (0001) ``C-cut'' and (10$\bar{1}$2)
``R-cut'' families of planes. It has space group $R \bar{3}c$ and can be
represented either by a rhombohedral unit cell with 10 atoms (2 Al$_2$O$_3$
molecular units) or a hexagonal cell with 30 atoms (6 Al$_2$O$_3$ molecular
units). In our calculations, we assigned the initial positions of the Al and
O ions according to the (experimental) values of Wyckoff.\cite{Wyckoff} In
the conventional hexagonal cell, there are 6 oxygen planes organised in the
$...ABABAB...$ closed-packed stacking sequence in the $c$ direction, with the
aluminum ions occupying 2/3 of the octahedral sites.

Apart from the $\alpha$ phase, there are two metastable polymorphs of alumina
which are of practical interest, viz.\ $\kappa$-Al$_2$O$_3$ and
$\gamma-{\rm{}Al}_2{\rm{}O}_3$; we focus on $\kappa$-Al$_2$O$_3$ hereafter.
This polymorph can be synthesized by CVD and, because of its hardness, is
used as surface coating.\cite{Ruppi} The crystalline structure of
$\kappa$-Al$_2$O$_3$ is orthorhombic, space group $Pna2_1$, with 40 atoms (8
Al$_2$O$_3$ molecular units) in the unit cell.\cite{Liu2} There are four
oxygen planes in the cell, organized in the $...ABCABCABC...$ stacking
sequence, and the aluminum ions fill the tetrahedral and/or octahedral
sites.\cite{Holm} The positions of the Al atoms in this structure is under
debate. Indeed, Belonoshko {\em et al}.\cite{Belonoshko} proposed a model in
which $2/3$ of the octahedral sites are filled with Al ions as in
$\alpha$-Al$_2$O$_3$ whereas, according to the theoretical studies of
Yourdshahyan {\em et al}., the most stable structure is one of the nine
possible configurations for which the Al are only in octahedral
positions.\cite{Yourdshahyan1, Yourdshahyan2} In this work, for the sake of
simplicity and clarity, we have chosen to fix the reduced coordinates of Al
and O ions to the experimental values issued from the Rietveld refinement of
Smrcok and al.;\cite{Smrcok} thus, all the ions are in position $4a$ with
coordinates $(x_1,y_1,z_1)$ for Al and $(x_2,y_2,z_2)$ for O, with symmetry
operations $(x,y,z;-x,-y,z+1/2;x+1/2,-y+1/2,z;-x+1/2,y+1/2,z+1/2)$.

The $E(V)$ curves for both polymorphs are presented in Fig.\
\ref{Fig_E_vs_V_al2o3} and the parameters of the Murnaghan equation of state
are provided in Table \ref{Tab_Murnaghan_1}. One can see that our {\em ab
initio} calculations do reproduce the correct relative stability of the two
phases; it is indeed known that at low pressure the
$\kappa$-${\rm{}Al}_2{\rm{}O}_3 \rightarrow \alpha$-${\rm{}Al}_2{\rm{}O}_3$
phase transition occurs around 1000\textsuperscript{o}C.

\begin{table}[tb]
\caption{Parameters of the Murnaghan equation of state, $V_0$
(\AA\textsuperscript{3}/atom), $B_0$ (GPa), and $B'_0$, for
$\alpha$-Al$_2$O$_3$ and $\kappa$-Al$_2$O$_3$, and comparison with other
results from the literature.}
\begin{ruledtabular}
\begin{tabular}{lccccccccccccc}
                   & \multicolumn{6}{c}{$\alpha$-Al$_2$O$_3$} &  & \multicolumn{6}{c}{$\kappa$-Al$_2$O$_3$} \\
                   \cline{2-7} \cline{9-14}
                   & \multicolumn{2}{c}{$V_0$} & \multicolumn{2}{c}{$B_0$} & \multicolumn{2}{c}{$B^{'}_0$} &  & \multicolumn{2}{c}{$V_0$} & \multicolumn{2}{c}{$B_0$} & \multicolumn{2}{c}{$B^{'}_0$} \\ \hline
LDA \footnote{This work} & \multicolumn{2}{c}{8.38} &  \multicolumn{2}{c}{260} &  \multicolumn{2}{c}{4} &  & \multicolumn{2}{c}{8.84} &  \multicolumn{2}{c}{239} &  \multicolumn{2}{c}{4.3} \\
LDA \footnote{Ref.\ \onlinecite{Siegel}} & \multicolumn{2}{c}{ } &  \multicolumn{2}{c}{239} &  \multicolumn{2}{c}{ } &  & \multicolumn{2}{c}{} &  \multicolumn{2}{c}{} &  \multicolumn{2}{c}{} \\
LDA \footnote{Ref.\ \onlinecite{Boettger}} & \multicolumn{2}{c}{8.51} &  \multicolumn{2}{c}{244} &  \multicolumn{2}{c}{4.305} &  & \multicolumn{2}{c}{} &  \multicolumn{2}{c}{} &  \multicolumn{2}{c}{} \\
LDA \footnote{Ref.\ \onlinecite{Marton}} & \multicolumn{2}{c}{8.36} &  \multicolumn{2}{c}{257} &  \multicolumn{2}{c}{4.05} &  & \multicolumn{2}{c}{} &  \multicolumn{2}{c}{} &  \multicolumn{2}{c}{} \\
Exp. \footnote{Ref.\ \onlinecite{Amour}} & \multicolumn{2}{c}{8.53} &  \multicolumn{2}{c}{254.4} &  \multicolumn{2}{c}{4.275} &  & \multicolumn{2}{c}{} &  \multicolumn{2}{c}{} &  \multicolumn{2}{c}{} \\
Emp. \footnote{Ref.\ \onlinecite{Belonoshko}; `Emp.' stands for `empirical model'} & \multicolumn{2}{c}{} &  \multicolumn{2}{c}{} &  \multicolumn{2}{c}{} &  & \multicolumn{2}{c}{8.802} &  \multicolumn{2}{c}{229.2} &  \multicolumn{2}{c}{ } \\
LDA \footnote{Ref.\ \onlinecite{Yourdshahyan2}} & \multicolumn{2}{c}{} &  \multicolumn{2}{c}{} &  \multicolumn{2}{c}{} &  & \multicolumn{2}{c}{8.754} &  \multicolumn{2}{c}{251.8} &  \multicolumn{2}{c}{ } \\
\end{tabular}
\end{ruledtabular}
\label{Tab_Murnaghan_1}
\end{table}

The lattice parameters of the rhombohedral and hexagonal phases of
$\alpha$-Al$_2$O$_3$ are summarized in Table \ref{Tab_latt_param_1}. The
parameters for the hexagonal cell, $\vec{a}_{\rm h1}$, $\vec{a}_{\rm h2}$,
and $\vec{c}_{\rm h}$, are deduced from the rhombohedral ones, $\vec{a}_{\rm
rh1}$, $\vec{a}_{\rm rh2}$, and $\vec{a}_{\rm rh3}$, as
follows:\cite{Cousins}
   \begin{eqnarray}
   \vec{a}_{\rm h1} & = & \vec{a}_{\rm rh1}-\vec{a}_{\rm rh2}  \\
   \vec{a}_{\rm h2} & = & \vec{a}_{\rm rh2}-\vec{a}_{\rm rh3}  \\
   \vec{c}_{\rm h} & = & \vec{a}_{\rm rh1}+\vec{a}_{\rm rh2}+\vec{a}_{\rm rh3}
   \end{eqnarray}
with $a_{\rm rh}=||\vec{a}_{\rm rh1}||=||\vec{a}_{\rm rh2}||=||\vec{a}_{\rm
rh3}||$, $a_h=||\vec{a}_{\rm h1}||=||\vec{a}_{\rm h2}||$, and
$c_h=||\vec{c}_{\rm h}||$. The parameters for (orthorhombic)
$\kappa$-Al$_2$O$_3$ are $a_{0}$, $b_{0}$ and $c_{0}$.

\begin{table}[tb]
\caption{Lattice parameters for $\alpha$-Al$_2$O$_3$ in the rhombohedral and
hexagonal cells, and for $\kappa$-Al$_2$O$_3$ in the orthorhombic cell
(\AA), and comparison with other results from the literature; $\alpha_{\rm rh}$
is the angle of the rhombohedral cell (degrees).}
\begin{ruledtabular}
\begin{tabular}{lcccccccc}
                  & \multicolumn{4}{c}{$\alpha$-Al$_2$O$_3$} &  & \multicolumn{3}{c}{$\kappa$-Al$_2$O$_3$} \\
                  \cline{2-5} \cline{7-9}
                  & $a_{\rm rh}$ & $\alpha_{\rm rh}$ & $a_h$ & $c_h$ &  & $a_{0}$ & $b_{0}/a_{0}$ & $c_{0}/a_{0}$ \\ \hline
LDA \footnote{This work}    & 5.114 & 54.952 & 4.717 & 12.987 &  & 4.836 & 1.711 & 1.835 \\
LDA \footnote{Ref.\ \onlinecite{Siegel}} &  &  & 4.714 & 12.861 &  &  &  &  \\
LDA \footnote{Ref.\ \onlinecite{Boettger}} &  &  & 4.767 & 12.969 &  &  &  &  \\
Emp. \footnote{Ref.\ \onlinecite{Sun2}} &  &  & 4.773 & 12.990 &  &  &  &  \\
Exp. \footnote{Ref.\ \onlinecite{Cousins}} &  &  & 4.7589 & 12.991 &  &  &  &  \\
Exp. \footnote{Ref.\ \onlinecite{Lewis}} &  &  & 4.760 & 12.993 &  &  &  &  \\
Exp. \footnote{Ref.\ \onlinecite{Wyckoff}} & 5.128 & 55.333 & 4.7628 & 13.0032 &  &  &  &  \\
Emp. \footnote{Ref.\ \onlinecite{Belonoshko}} &  &  &  &  &  & 4.770 & 1.731 & 1.874 \\
LDA \footnote{Ref.\ \onlinecite{Yourdshahyan2}} &  &  &  &  &  & 4.804 & 1.7137 & 1.8435 \\
Exp. \footnote{Ref.\ \onlinecite{Smrcok}} &  &  &  &  &  & 4.8340 & 1.719 & 1.8480 \\
Exp. \footnote{Ref.\ \onlinecite{Liu2}} &  &  &  &  &  & 4.69 & 1.744 & 1.891 \\
\end{tabular}
\end{ruledtabular}
\label{Tab_latt_param_1}
\end{table}

\begin{figure}[tb]
\includegraphics[width=8cm]{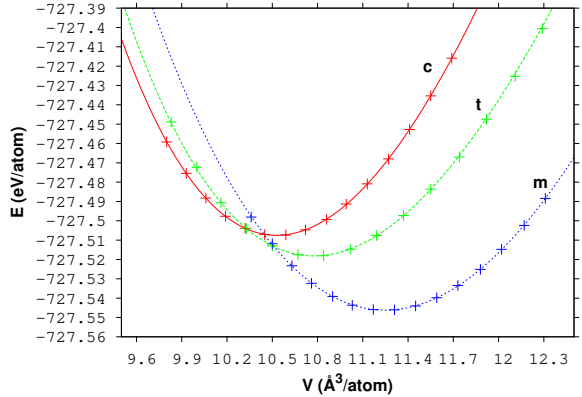}
\caption{(Color online) Energy vs volume for the low-pressure polymorphs of
ZrO$_2$, as indicated}
\label{Fig_E_vs_V_zro2}
\end{figure}

We now turn to zirconium dioxide, an important material for applications in
optical, mechanical and thermal coatings. Here we are concerned with the
low-pressure polymorphs of ZrO$_2$ stoichiometry. From 0 to 1400 K, ZrO$_2$
is monoclinic (`m'; this phase is called baddeleyite), of space group
$P2_1/c$;\cite{McCullough} between 1400 and 2650 K, it is tetragonal (`t'),
of space group $P4_2/nmc$;\cite{Teufer} finally, above 2650 K and all the way
to the melting point, it is cubic (`c'), of space group
$Fm\bar{3}m$.\cite{Howard} Here we describe the three phases through their
conventional unit cells with 12 atoms (4 ZrO$_2$ molecular units). Table
\ref{Tab_Murnaghan_2} presents the parameters of the Murnaghan equation of
state derived from the $E(V)$ curves of Fig.\ \ref{Fig_E_vs_V_zro2}. Our
calculations reproduce the correct relative stability of the three polymorphs
and our fitted parameters agree with previous theoretical and experimental
results. The structural parameters are provided in Table
\ref{Tab_latt_param_2}.

\begin{table}[tb]
\caption{Parameters of the Murnaghan equation of state, $V_0$
(\AA\textsuperscript{3}/atom), $B_0$ (GPa), and $B'_0$, for m-, t-, and
c-ZrO$_2$, and comparison with other results from the literature.}
\begin{ruledtabular}
\begin{tabular}{lccccccccccccc}
            & \multicolumn{6}{c}{LDA \footnote{This work}} &  & \multicolumn{6}{c}{LDA \footnote{Ref.\ \onlinecite{Stapper}}} \\
\cline{2-7} \cline{9-14}
& \multicolumn{2}{c}{m} & \multicolumn{2}{c}{t} & \multicolumn{2}{c}{c} &  & \multicolumn{2}{c}{m} & \multicolumn{2}{c}{t} & \multicolumn{2}{c}{c} \\ \hline
$V_0$ & \multicolumn{2}{c}{11.25}  & \multicolumn{2}{c}{10.78} &  \multicolumn{2}{c}{10.53}  &  & \multicolumn{2}{c}{11.68}  & \multicolumn{2}{c}{11.13} & \multicolumn{2}{c}{10.91} \\
$B_0$ & \multicolumn{2}{c}{203} & \multicolumn{2}{c}{225} &  \multicolumn{2}{c}{273} &  & \multicolumn{2}{c}{185} & \multicolumn{2}{c}{197} & \multicolumn{2}{c}{268} \\
$B^{'}_0$ & \multicolumn{2}{c}{2.4} & \multicolumn{2}{c}{4.7} &  \multicolumn{2}{c}{4.3} &  & \multicolumn{2}{c}{1.8} & \multicolumn{2}{c}{5.0} & \multicolumn{2}{c}{3.6} \\
\hline
            & \multicolumn{6}{c}{Experiment} \\
\cline{2-7}
& \multicolumn{2}{c}{m} & \multicolumn{2}{c}{t} & \multicolumn{2}{c}{c} \\ \hline
$V_0$ & \multicolumn{2}{c}{11.74 \footnote{Ref.\ \onlinecite{Howard}}} & \multicolumn{2}{c}{11.64 \footnote{Ref.\ \onlinecite{Teufer}}} &  \multicolumn{2}{c}{10.86 \footnote{Ref.\ \onlinecite{Wyckoff}}} \\
$B_0$ & \multicolumn{2}{c}{95-185 \footnote{Ref.\ \onlinecite{Leger}}} & \multicolumn{2}{c}{190-185 \footnote{Refs.\ \onlinecite{Fukuhara1}, \onlinecite{Fukuhara2}}} & \multicolumn{2}{c}{194-220 \footnote{Refs.\ \onlinecite{Aldebert}, \onlinecite{Kandil}, \onlinecite{Ingel}} 190 \footnote{Ref.\ \onlinecite{Botha}}} \\
$B^{'}_0$ & \multicolumn{2}{c}{4-5 \textsuperscript{\textit{a}}} & \multicolumn{2}{c}{ } &  \multicolumn{2}{c}{ } \\
\end{tabular}
\end{ruledtabular}
\label{Tab_Murnaghan_2}
\end{table}

\begin{table}[tb]
\caption{Lattice parameters for m-, t-, and c-ZrO$_2$ (\AA), and comparison
with other results from the literature; $\beta$ is the angle of the
monoclinic phase (degrees) and $d_z$ is the tetragonal distorsion of $O$
atoms in the $\vec{c}$ direction of the tetragonal phase.}
\begin{ruledtabular}
\begin{tabular}{lcccc}
              & \multicolumn{4}{c}{m} \\
              \cline{2-5}
              & $a$ & $b/a$ & $c/a$ & $\beta$ \\ \hline
LDA \footnote{This work} & 5.085 & 1.020 & 1.025 & 99.31 \\
LSDA \footnote{Ref.\ \onlinecite{Christensen3}} & 5.136 & 1.020 & 1.029 & 99.43 \\
Exp. \footnote{Refs.\ \onlinecite{Howard} for m, \onlinecite{Teufer} for t, and \onlinecite{Wyckoff} for c} & 5.1505  & 1.0119 & 1.0317 & 99.230 \\
              & \multicolumn{3}{c}{t} & c \\
              \cline{2-4} \cline{5-5}
              & $a$ & $c/a$ & $d_z$ & $a$ \\ \hline
LDA \textsuperscript{\textit{a}} & 5.046 & 1.029 & 0.049 & 5.02 \\
LSDA \textsuperscript{\textit{b}} & 5.086 & 1.013 & 0.040 & 5.082 \\
Exp. \textsuperscript{\textit{c}} & 5.15 & 1.02 & 0.065 & 5.07 \\
\end{tabular}
\end{ruledtabular}
\label{Tab_latt_param_2}
\end{table}

Finally, we discuss the structural parameters of Y$_2$O$_3$ (bixbyite), which
is body-centered cubic, space group $Ia\bar{3}$. This can be viewed in the
conventional unit cell with 80 atoms and lattice parameter $a=10.604$
\AA,\cite{Wyckoff} or in the primitive unit cell with 40 atoms and lattice
parameter $a'=a \sqrt{3}/2$. In this work, we have used the primitive unit
cell; there are two nonequivalent yttrium sites, $8a$ for Y$_I$ of
coordinates ($1/4,1/4,1/4$) and $24d$ for Y$_{II}$ of coordinates
($u,0,1/4$). The oxygen site is $48e$ of coordinates ($x,y,z$). The relevant
symmetry operations can be found elsewhere.\cite{Wyckoff} Table
\ref{Tab_latt_param_3} presents the parameters of the Murnaghan equation of
state, the lattice parameter, and the internal ionic positions for Y$_{II}$
and O.

\begin{table}[tb]
\caption{Lattice parameter $a$ (\AA), parameters of the Murnaghan equation of
state, $V_0$ (\AA\textsuperscript{3}/atom), $B_0$ (GPa) and $B^{'}_0$, and
Wyckoff coordinates of $Y_{II}$ and O for the Y$_2$O$_3$ bixbyite structure,
and comparison with other results from the literature.}
\begin{ruledtabular}
\begin{tabular}{lccccccccccccc}
              &     & \multicolumn{2}{c}{} & \multicolumn{2}{c}{} & \multicolumn{2}{c}{} & \multicolumn{2}{c}{$Y_{II}$} & \multicolumn{4}{c}{$O$} \\
              & $a$ & \multicolumn{2}{c}{$V_0$} & \multicolumn{2}{c}{$B_0$} & \multicolumn{2}{c}{$B^{'}_0$} & \multicolumn{2}{c}{$-u$} & \multicolumn{4}{c}{($x,y,z$)} \\ \hline
LDA \footnote{This work} & 10.481 & \multicolumn{2}{c}{14.31} & \multicolumn{2}{c}{160} & \multicolumn{2}{c}{4.4} &\multicolumn{2}{c}{0.0326} & \multicolumn{4}{c}{(0.3904,0.1512,0.3798)} \\
LDA \footnote{Ref.\ \onlinecite{Stapper}} & 10.483 & \multicolumn{2}{c}{} & \multicolumn{2}{c}{143} & \multicolumn{2}{c}{3.9} &\multicolumn{2}{c}{0.0327} & \multicolumn{4}{c}{(0.3905,0.1518,0.3803)} \\
Exp. \footnote{Ref.\ \onlinecite{Wyckoff}} & 10.604 & \multicolumn{2}{c}{} & \multicolumn{2}{c}{} & \multicolumn{2}{c}{} &\multicolumn{2}{c}{0.0314} & \multicolumn{4}{c}{(0.3890,0.1500,0.3770)} \\
\end{tabular}
\end{ruledtabular}
\label{Tab_latt_param_3}
\end{table}

The above results clearly establish the ability of the PAW method to
reproduce the correct structural properties of the systems we are concerned
with. In what follows, we present first the models used to simulate the
(0001) surface of $\alpha$-Al$_2$O$_3$ as well as the (100) and (111)
surfaces of YSZ. We then discuss the method for constructing the interface
supercells, which will be used to calculate the interface energies.


\section{Interface energies of $\{$${\rm YSZ} ||
(0001)_{\alpha-{\rm{}Al}_2{\rm{}O}_3}$$\}$ (1), (2) and (3)}
\label{unrelaxed}

In order to calculate the interface energies, we constructed supercell models
consisting of a $(0001)_{\alpha-{\rm{}Al}_2{\rm{}O}_3}$ slab for the
substrate and a $(100)_{\rm YSZ}$ or $(111)_{\rm YSZ}$ slab for the epitaxial
layer. The supercells are parallelepipeds, and the interface is taken to be
perpendicular to the $z$ direction. Before proceeding, however, we consider
the free surfaces and compute their unrelaxed and relaxed energies. Periodic
boundary conditions are used; for free surfaces, a vacuum region is inserted
in the supercell. Thus, in all cases there are two interfaces, either between
the two materials or between the surface of the material and the vacuum.

\begin{table}[tb]
\caption{Free surface energies $\gamma_{(0001)}$ (J/m\textsuperscript{2}) of
the unrelaxed and relaxed $\alpha$-Al$_2$O$_3$ models for $N=9,12,15,18$
atomic layers (i.e., $\Delta N=3$ here).}
\begin{ruledtabular}
\begin{tabular}{cccccc}
$N$     & Unrelaxed \footnote{This work} &  & Relaxed \textsuperscript{\textit{a}} &  & Relaxed \footnote{Ref.\ \onlinecite{Siegel}} \\
\cline{2-2} \cline{4-4} \cline{6-6}
9       & 4.14 &  & 2.08 &  & 2.02 \\
12      & 4.22 &  & 2.14 &  &      \\
15      & 4.26 &  & 2.12 &  & 2.12 \\
18      & 4.26 &  & 2.12 &  &      \\
21      &      &  &      &  & 2.12 \\
27      &      &  &      &  & 2.12 \\
\end{tabular}
\end{ruledtabular}
\label{Tab_results_1}
\end{table}

The thickness of the slab must be sufficient to yield converged results and
yet remain computationally manageable. To this end, one may first define two
structures --- one for the bulk and one for the slab --- having the same
number of atomic layers $N$; the slab has two free surfaces owing to the
presence of a vacuum region (see above). The free surface energy is then
defined as the excess energy of the slab [$(hkl)$ indices] relative to the
bulk divided by the surface $S$ of the slab:
   \begin{eqnarray}
   \gamma_{hkl}(N) & = & \frac{E_{\rm slab}(N)-\frac{N}{\Delta N} \Delta E_{\rm bulk}(N)}{S} \label{eq9} \\
   \Delta E_{\rm bulk}(N) & = & E_{\rm bulk}(N)-E_{\rm bulk}(N- \Delta N),
   \end{eqnarray}
where $\Delta N$ is the difference in the number of layers between two
different slab models. The convergence of $\gamma_{hkl}$ can be studied as a
function of $N$.

However, Boettger has shown that this approach is not very
accurate,\cite{Boettger1} as the convergence of $\gamma_{hkl}(N)$ depends on
the thickness of both the slab and the bulk, but also on $\Delta E_{\rm
bulk}(N)$. He proposed to use, instead:
   \begin{eqnarray}
   \gamma_{hkl}(N) & = & \frac{E_{\rm slab}(N)-\frac{N}{\Delta N} \Delta E_{\rm slab}(N')}{S} \label{eq12} \\
   \Delta E_{\rm slab}(N') & = & E_{\rm slab}(N')-E_{\rm slab}(N'-\Delta N)
   \end{eqnarray}
where $N'$ is the number of layers for which the value of $\Delta E_{\rm
slab}$ is sufficiently converged to ensure that the behaviour of
$\gamma_{hkl}$ is a function of $N$ only. Thus, in this approach, the term
$\Delta E_{\rm bulk}(N)$ in Eq.\ \ref{eq9} is replaced by $\Delta E_{\rm
slab}(N')$ in Eq.\ \ref{eq12}, thereby reducing the convergence study of
$\gamma_{hkl}$ to slab calculations. For the sake of comparison with previous
{\em ab initio} results on this system,\cite{Siegel} we used the Boettger
method here.

\subsection{$(0001)_{\alpha-{\rm{}Al}_2{\rm{}O}_3}$ free surface energy}
\label{free1}

Several calculations of stoichiometric and non stoichiometric free surface
energies of $(0001)_{\alpha-{\rm{}Al}_2{\rm{}O}_3}$ have been reported in the
literature.\cite{Guo, Godin, Batyrev2, Tepesch, Marmier, Sun1, Sun2} Here we
deal only with stoichiometric systems; this choice is not restrictive as it
was demonstrated that the most stable (0001) surface of $\alpha$-Al$_2$O$_3$
is stoichiometric, Al-terminated, in a wide range of $P_{{\rm
O}_2}$.\cite{Batyrev2, Wang} Further, previous {\em ab initio}
calculations\cite{Siegel} have shown that a vacuum thickness of 10 \AA\ is
sufficient and this is the value we have used. We have nevertheless studied
the convergence with regard to the thickness of the solid, viz.\ 9, 12, 15,
and 18 atomic layers.

The results, presented in Table \ref{Tab_results_1}, are found to be in very
good agreement with those of Siegel.\cite{Siegel} One may note the huge
differences between the unrelaxed and relaxed energies --- the absolute
differences are $\sim$2 J/m\textsuperscript{2}. This is a consequence of the
inward relaxation of the atomic planes in the $z$ direction. Table
\ref{Tab_results_2} gives the average relaxation of the atomic planes
relative to the original bulk spacing. As found in previous calculations, the
inward relaxation of the Al atomic plane is close to 80\% and leads to the
formation of $sp^2$-like atomic bonding at the free surface. The inward
relaxation of the top Al plane is related to the increase of the electronic
density, yielding a lower free surface energy. Our study demonstrates that 15
atomic layers are needed to model accurately the bulk structure, but the
results are already quite reasonable for $N=9$, offering a good compromise
between accuracy and computational workload as we discuss in Sec.\
\ref{model_interfaces}.

\begin{table}[tb]
\caption{Average relaxation of the atomic planes in the [001] direction for
$\alpha$-Al$_2$O$_3$, expressed as a proportion of the initial bulk spacing
for $N=9,12,15,18$ atomic layers.}
\begin{ruledtabular}
\begin{tabular}{lccccccccccc}
      & \multicolumn{8}{c}{LDA \footnote{This work}} &  & \multicolumn{2}{c}{LDA \footnote{Ref.\ \onlinecite{Siegel}}} \\
\cline{2-9} \cline{11-12}
$N$  & \multicolumn{2}{c}{9} & \multicolumn{2}{c}{12} & \multicolumn{2}{c}{15} & \multicolumn{2}{c}{18} &  & \multicolumn{2}{c}{15} \\
\cline{1-12}
Al-O & \multicolumn{2}{c}{$-$87} & \multicolumn{2}{c}{$-$84} & \multicolumn{2}{c}{$-$83} & \multicolumn{2}{c}{$-$83} &  & \multicolumn{2}{c}{$-$83} \\
O-Al & \multicolumn{2}{c}{5} & \multicolumn{2}{c}{5} & \multicolumn{2}{c}{5} & \multicolumn{2}{c}{5} &  & \multicolumn{2}{c}{3} \\
Al-Al & \multicolumn{2}{c}{$-$52} & \multicolumn{2}{c}{$-$44} & \multicolumn{2}{c}{$-$44} & \multicolumn{2}{c}{$-$46} &  & \multicolumn{2}{c}{$-$46} \\
Al-O & \multicolumn{2}{c}{23} & \multicolumn{2}{c}{20} & \multicolumn{2}{c}{18} & \multicolumn{2}{c}{19} &  & \multicolumn{2}{c}{19} \\
O-Al & \multicolumn{2}{c}{23} & \multicolumn{2}{c}{6} & \multicolumn{2}{c}{5} & \multicolumn{2}{c}{4} &  & \multicolumn{2}{c}{4} \\
\end{tabular}
\end{ruledtabular}
\label{Tab_results_2}
\end{table}

\subsection{$(100)_{\rm YSZ}$ and $(111)_{\rm YSZ}$ free surface energies}
\label{free_2}

We now consider the YSZ (100) and (111) free surface energies. We follow the
approach proposed by Stapper {\em et al}.\cite{Stapper} and Ballabio {\em et
al}.\cite{Ballabio} to build the bulk and slab structures. YSZ is a solid
solution of Y$_2$O$_3$ in ZrO$_2$, of space group $Fm\bar{3}m$ (as
c-ZrO$_2$). Proper simulation of YSZ depends on two parameters: (i) The size
of the simulation cell, which must be large enough to provide a good
statistical representation of the proportion of Y atoms and O vacancies
($V_{\rm O}$) for a given molar proportion of Y$_2$O$_3$. (ii) The ground
state energy, which depends on the relative positions of Y ions and O
vacancies, and which cannot be chosen at random: Stapper {\em et al}.\ have
indeed shown that the most stable configuration is that for which the O
vacancies are next-nearest neighbours to yttrium atoms.\cite{Stapper}

Here, the doping level of Y$_2$O$_3$ is set to 10\% molar, consistent with
the experimental studies of Bachelet {\em et al}.\cite{Bachelet} For
consistency and comparison with previous works, the positions of the Y ions
and the O vacancies are chosen such that two $V_{\rm O}$'s cannot be closer
to one another than third nearest neighbour; two Y's can be nearest
neighbours but a Y cannot be closer to a $V_{\rm O}$ than next nearest
neighbour. The bulk cell of $(100)_{\rm YSZ}$ is made up of 4 (Zr,Y) and 5
(O,$V_{\rm O}$) atomic layers in the [100] direction ($N=9$), for a total of
93 atoms (26 Zr, 61 O, 6 Y; 3 $V_{\rm O}$). The dimensions of the bulk cell
are $\delta x = \delta y = \delta z = 2a_{\rm YSZ}$, where the theoretical
lattice parameter is derived from the experimental relation established by
Pascual and D\'uran,\cite{Pascual} $a_{\rm YSZ} = a_0 + 0.003x$, with $x$ the
molar percent of Y$_2$O$_3$. Using the values of $a_0$ given in Table
\ref{Tab_latt_param_2} and setting $x=0.1$, we obtain $a_{\rm YSZ}=5.05$ \AA.
For the $(111)_{\rm YSZ}$ cell, we have 3 (Zr,Y) and 6 (O,$V_{\rm O}$) atomic
layers in the [111] direction ($N=9$), for a total of 140 atoms (40 Zr, 92 O,
8 Y; 4 $V_{\rm O}$). The dimensions of the cell are $ \delta x =
2\sqrt{2}a_{\rm YSZ}$, $\delta y = 2\sqrt{3/2}a_{\rm YSZ}$, and $\delta z
\sqrt{3}a_{\rm YSZ}$.

From these bulk cells, surface slabs are constructed by introducing a 10
\AA-thick vacuum layer along $z$. For consistency with the case of
(Al-terminated) $\alpha$-Al$_2$O$_3$, we also consider stoichiometric
surfaces hereafter so that, in both cases, the cells are terminated by an
oxygen plane. In the case of $(100)_{\rm YSZ}$, this requires half of the O
atoms to be removed from each side of the slab. We placed one oxygen vacancy
on each side of the slab cells on the free surfaces. The (100) and (111) free
surface energies of YSZ, computed using Eq.\ \ref{eq9}, are presented in
Table \ref{Tab_results_3}; the areas of the free surfaces are $S=8a_{\rm
YSZ}^2$ and $S=8\sqrt{3}a_{\rm YSZ}^2$, respectively.

\begin{table}[tb]
\caption{Unrelaxed and relaxed stoichiometric free surface energies
$\gamma_{(100)}$ and $\gamma_{(111)}$ for YSZ (J/m\textsuperscript{2}).}
\begin{ruledtabular}
\begin{tabular}{ccccccccc}
\multicolumn{4}{c}{Unrelaxed} &  & \multicolumn{4}{c}{Relaxed} \\
\cline{1-4} \cline{6-9}
\multicolumn{2}{c}{$\gamma_{(100)}$} & \multicolumn{2}{c}{$\gamma_{(111)}$} &  & \multicolumn{2}{c}{$\gamma_{(100)}$} & \multicolumn{2}{c}{$\gamma_{(111)}$} \\
\cline{1-2} \cline{3-4} \cline{6-7} \cline{8-9}
\multicolumn{2}{c}{2.79 \footnote{This work}} & \multicolumn{2}{c}{1.30 \textsuperscript{\textit{a}}} &  & \multicolumn{2}{c}{1.71 \textsuperscript{\textit{a}}} & \multicolumn{2}{c}{1.17 \textsuperscript{\textit{a}}} \\
\multicolumn{2}{c}{} & \multicolumn{2}{c}{} &  & \multicolumn{2}{c}{1.75 \footnote{Ref.\ \onlinecite{Ballabio}}} & \multicolumn{2}{c}{1.04 \textsuperscript{\textit{b}}} \\
\end{tabular}
\end{ruledtabular}
\label{Tab_results_3}
\end{table}

We find good agreement with Ballabio {\em et al}.\ for the relaxed value of
$\gamma_{(100)}$, but there is a small difference for $\gamma_{(111)}$; this
might be due to the use of different relaxation schemes --- BFGS here, vs
Car-Parinello molecular dynamics\cite{Car} in Ref.\ \onlinecite{Ballabio}.
More important, however, we observe large changes arising from relaxation:
$\sim$1.2 J/m\textsuperscript{2} for $\gamma_{(100)}$ and $\sim$0.2
J/m\textsuperscript{2} for $\gamma_{(111)}$, with an average inward
relaxation of the top O plane of $\sim$25\% and $\sim$8\%, respectively.
These results are not surprising since (111) corresponds to a dense
arrangement of the atomic planes, which is not the case for (100).

\subsection{$\{$${\rm YSZ} || (0001)_{\alpha-{\rm{}Al}_2{\rm{}O}_3}$$\}$
interfaces}
\label{model_interfaces}

We now turn to the (1), (2), and (3) $\{$${\rm YSZ} ||
(0001)_{\alpha-{\rm{}Al}_2{\rm{}O}_3}$$\}$ interface models, constructed from
the structures discussed in the previous sections. More specifically, several
unit cells must be assembled in the ($x$,$y$) plane so as to minimize the
lattice mismatch; since the computational effort increases very rapidly with
size, the number of cells of each material must be chosen such that the
mismatch is no larger than a few percent in each supercell.

\begin{figure}[tb]
\includegraphics{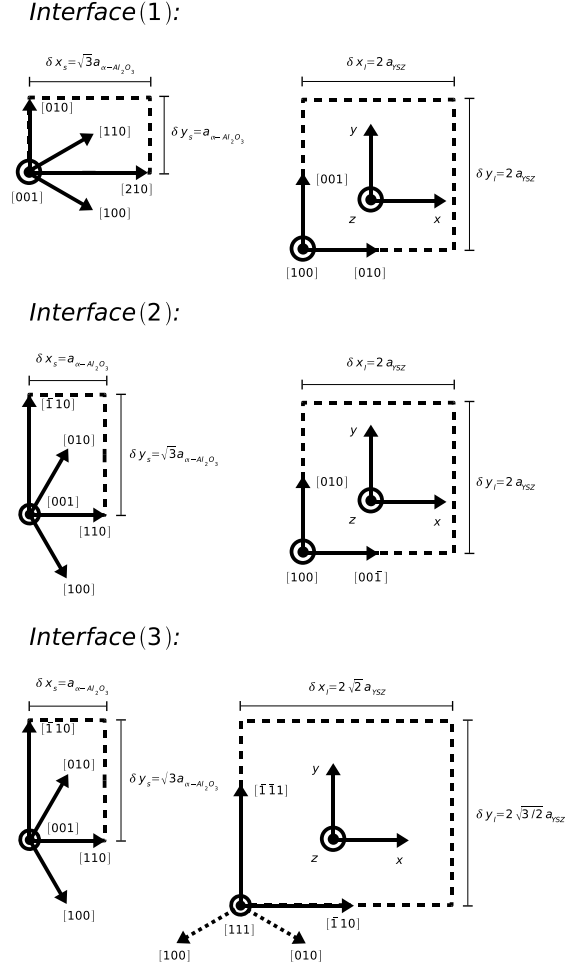}
\caption{Relative crystallographic orientations and dimensions of the
$\alpha$-Al$_2$O$_3$ and YSZ unit cells (dashed lines) in the $x$ and $y$
directions for each interface model. For model (3), the YSZ [100] and [010]
crystallographic orientations are out of the ($x$, $y$) plane and are
represented as dotted lines.}
\label{mismatch_1}
\end{figure}

\begin{figure*}
\includegraphics[scale=0.8]{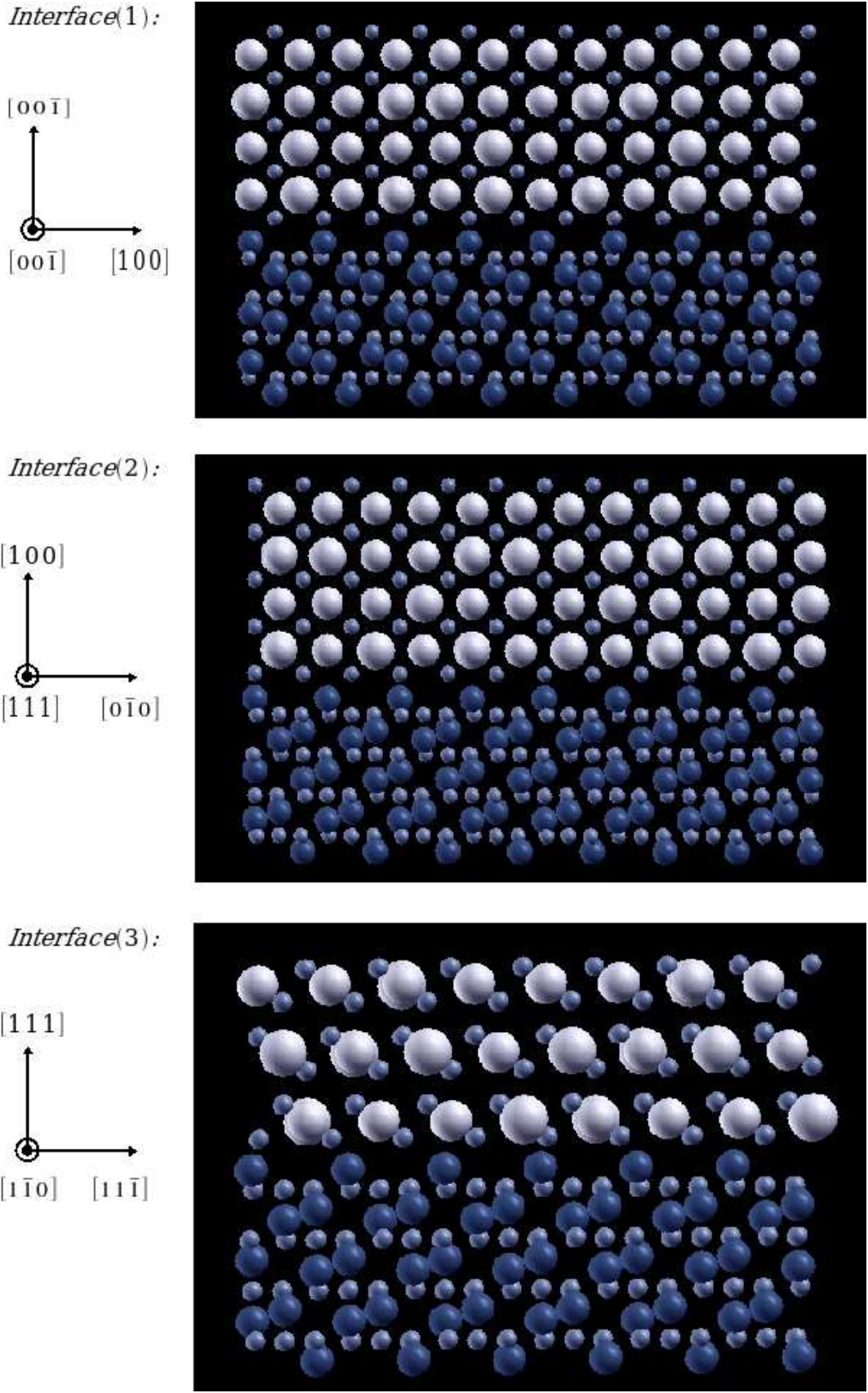}
\caption{(Color online) Side views of the unrelaxed interface models (1), (2)
and (3). The Y and Zr atoms are white, the O atoms are light blue and the Al
atoms are purple. The crystallographic orientations are those of the YSZ
phase.}
\label{models}
\end{figure*}

Figure \ref{mismatch_1} provides a schematic representation of the relative
orientations of the two materials; ball-and-stick models are presented in
Fig.\ \ref{models}. In practice, keeping the mismatch to within a few percent
would require unit cells containing at least 500 atoms. This is clearly
unmanageable. In order to reduce the workload, we fixed the thickness of the
$\alpha$-Al$_2$O$_3$ substrate to 9 atomic layers; as mentioned earlier,
while the system parameters are not fully converged at this value, they are
nevertheless adequately described. In addition, as we will be comparing
different structures with the same number of layers, systematic errors will
cancel out to a large extent. The computational workload remains
considerable, but was alleviated by running the calculations in parallel on
up to 252 processors.

The details of the geometry of the three systems are given in Table
\ref{Tab_results_4}: the mismatches are in all cases less than $\sim$4\%. The
distance between the substrate and the layer was fixed by assuming that the
YSZ O plane at the interface lies at the position where an
$\alpha$-Al$_2$O$_3$ O plane would have been in an infinite system.

\begin{table}[tb]
\caption{Lattice mismatches $\epsilon$ relative to the initial bulk spacings
(\%) and total number of atoms $n_{\rm a}$ for each interface model. $N_{lx}$
and $N_{sx}$, and $N_{ly}$ and $N_{sy}$, are the number of unit cells for the
layer and the substrate in the $x$ and $y$ directions, respectively.}
\begin{ruledtabular}
\begin{tabular}{cccccc}
\multirow{4}{*}{(1)} & \multicolumn{2}{c}{$N_{lx}=3$} & \multicolumn{2}{c}{$N_{ly}=1$} & \multirow{4}{*}{$n_{\rm a}=519$} \\
            & \multicolumn{2}{c}{$N_{sx}=4$} & \multicolumn{2}{c}{$N_{sy}=2$} & \\
            & \multicolumn{2}{c}{$\epsilon_{lx}=3.93$} & \multicolumn{2}{c}{$\epsilon_{ly}=-3.30$} & \\
            & \multicolumn{2}{c}{$\epsilon_{sx}=-3.64$} & \multicolumn{2}{c}{$\epsilon_{sy}=3.53$} & \\
\multicolumn{6}{c}{} \\
\multirow{4}{*}{(2)} & \multicolumn{2}{c}{$N_{lx}=1$} & \multicolumn{2}{c}{$N_{ly}=3$} & \multirow{4}{*}{$n_{\rm a}=519$} \\
            & \multicolumn{2}{c}{$N_{sx}=2$} & \multicolumn{2}{c}{$N_{sy}=4$} & \\
            & \multicolumn{2}{c}{$\epsilon_{lx}=-3.30$} & \multicolumn{2}{c}{$\epsilon_{ly}=3.93$} & \\
            & \multicolumn{2}{c}{$\epsilon_{sx}=3.53$} & \multicolumn{2}{c}{$\epsilon_{sy}=-3.64$} & \\
\multicolumn{6}{c}{} \\
\multirow{4}{*}{(3)} & \multicolumn{2}{c}{$N_{lx}=1$} & \multicolumn{2}{c}{$N_{ly}=2$} & \multirow{4}{*}{$n_{\rm a}=550$} \\
            & \multicolumn{2}{c}{$N_{sx}=3$} & \multicolumn{2}{c}{$N_{sy}=3$} & \\
            & \multicolumn{2}{c}{$\epsilon_{lx}=-0.45$} & \multicolumn{2}{c}{$\epsilon_{ly}=-0.46$} & \\
            & \multicolumn{2}{c}{$\epsilon_{sx}=0.46$} & \multicolumn{2}{c}{$\epsilon_{sy}=0.47$} & \\
\end{tabular}
\end{ruledtabular}
\label{Tab_results_4}
\end{table}

\begin{figure*}
\includegraphics[scale=0.6]{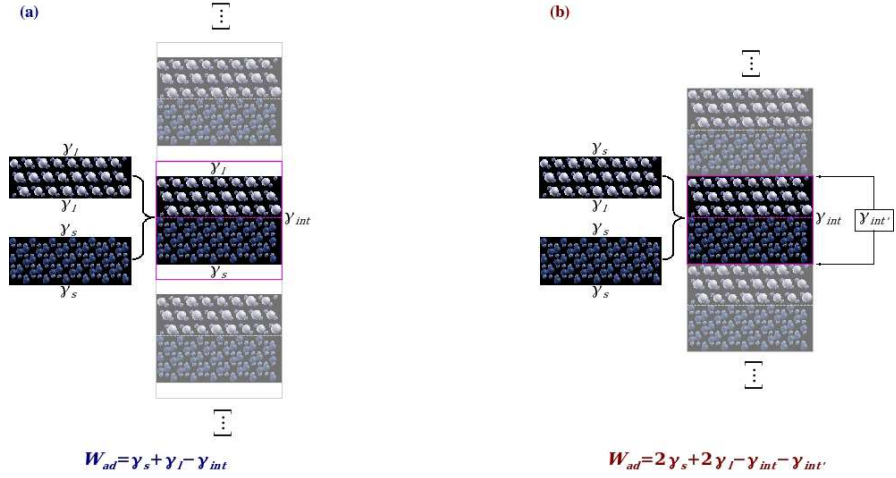}
\caption{(Color online) Schematic illustration of the geometry of the system and
corresponding interface energies: periodically-replicated model interface
structure (a) with vacuum and (b) without vacuum.}
\label{interface}
\end{figure*}

The total energy $E_{\rm int}$ of a given model structure can be related to
the ideal work of adhesion $W_{\rm ad}$ as follows:
   \begin{equation}\label{eq14}
   W_{\rm ad}=\frac{E_{s}+E_{l}-E_{\rm int}}{S},
   \end{equation}
where $E_{s}$ and $E_{l}$ are the total energies of the substrate and the
layer, respectively, and $S$ is the area of the interface. $W_{\rm ad}$ can be
expressed in terms of the interface and surface energies as:
   \begin{equation}\label{eq15}
   W_{\rm ad}=2\gamma_{s}+2\gamma_{l}-\gamma_{\rm int}-\gamma_{\rm int'},
   \end{equation}
\noindent
where $\gamma_{s}=\gamma_{\alpha-{\rm{}Al}_2{\rm{}O}_3}=\gamma_{(0001)}$ and
$\gamma_{l}=\gamma_{\rm YSZ}=\gamma_{\rm (100)}$ for interfaces (1) and (2)
$=\gamma_{\rm (111)}$ for interface (3). The quantity $\gamma_{\rm int'}$ is
the energy resulting from the presence of two interfaces in absence of
vacuum. The significance of the various quantities entering Eqs.\ \ref{eq14}
and \ref{eq15} in relation to the geometry of the models is schematically
illustrated in Fig.\ \ref{interface}.

One may argue that the interface energy between two solids is lower
than the sum of the free energies of the two surfaces:
   \begin{equation}\label{eq16}
   \gamma_{\rm int'} \le \gamma_{(0001)}+\gamma_{\rm YSZ}.
   \end{equation}
\noindent
Combining Eqs.\ \ref{eq14}, \ref{eq15}, and \ref{eq16} yields a lower bound
to the interface energy:
   \begin{equation}\label{eq17}
   \gamma_{\rm int} \ge \frac{E_{\rm int}-(E_{s}+E_{l})}{S}+\gamma_{(0001)}+\gamma_{\rm YSZ}.
   \end{equation}
\noindent
This is useful for comparing interface energies as $\gamma_{\rm YSZ}$ and
$\gamma_{(0001)}$ are known. However, because the mismatches are finite (cf.\
Table \ref{Tab_results_4}), the substrate and layers are both under strain
and, as a consequence, the values of $E_{s}$, $E_{l}$, $\gamma_{\rm YSZ}$,
and $\gamma_{(0001)}$ must reflect this. In practice, the values are adjusted
for the actual lattice parameters of the substrate and the layers; the
corresponding strained, unrelaxed free surface energies are listed in Table
\ref{Tab_results_5}.

\begin{table}[tb]
\caption{Free surface energies for the strained, unrelaxed systems (J/m$^2$).}
\begin{ruledtabular}
\begin{tabular}{cccccc}
        & Interface (1) &  & Interface (2) &  & Interface (3) \\
\cline{2-2} \cline{4-4} \cline{6-6}
        & $\gamma_{\rm (100)}$ &  & $\gamma_{\rm (100)}$ &  & $\gamma_{\rm (111)}$ \\
\cline{2-2} \cline{4-4} \cline{6-6}
        & 2.75 &  & 2.75 &  & 1.25 \\
$N$     & $\gamma_{(0001)}$ &  & $\gamma_{(0001)}$ &  & $\gamma_{(0001)}$ \\
\cline{2-2} \cline{4-4} \cline{6-6}
9       & 4.06 &  & 4.14 &  & 4.11 \\
12      & 4.12 &  & 4.20 &  & 4.20 \\
15      & 4.12 &  & 4.20 &  & 4.24 \\
18      & 4.12 &  & 4.20 &  & 4.24 \\
\end{tabular}
\end{ruledtabular}
\label{Tab_results_5}
\end{table}

One may note that for both layers and substrates, the strained unrelaxed free
surface energies are strictly lower than their corresponding unstrained
values (see Tables \ref{Tab_results_1} and \ref{Tab_results_3} for
comparison). This behaviour is related to both the evolution of the free
surface (contraction or dilatation, see Table \ref{Tab_results_4}) and the
modification of the electronic density. Using Eqs.\ \ref{eq16} and
\ref{eq17}, and the parameters listed in Table \ref{Tab_results_5}, we have
   \begin{eqnarray}
   4.48 \le & \gamma_{\rm int(1)} & \le 6.81, \\
   4.41 \le & \gamma_{\rm int(2)} & \le 6.89, \\
   2.46 \le & \gamma_{\rm int(3)} & \le 5.36.
   \end{eqnarray}
Our calculations demonstrate, therefore, that the interface energy for model
(3) tends to be lower than that of models (1) and (2), implying that the
fomer is the most thermodynamically stable of the three. This result is
consistent with the continuity of the three-fold
$\alpha-{\rm{}Al}_2{\rm{}O}_3$ (0001) symmetry axis with the YSZ (111)
symmetry axis in model (3),\cite{Bachelet} i.e., electronic bonding is much
stronger here than in models (1) and (2). As a consequence, on a perfect
$\alpha-{\rm{}Al}_2{\rm{}O}_3$ substrate, the creation of a type-(3)
interface requires much more energy than interfaces of types (1) and (2)
because it is more strongly bonded. Thus, at the beginning of the islanding
process, when the thin solid film still exhibits a large interface area with
the substrate, the formation of interfaces (1) and (2) is favored over
interface (3). As a consequence, $(111)_{\rm YSZ}$ islands are not expected
to form on a perfect $\alpha-{\rm{}Al}_2{\rm{}O}_3$ substrate --- only
$(100)_{\rm YSZ}$ islands with large interface areas should be observed.

On an imperfect substrate, now, the islands nucleating at the location of the
defects are subject to enhanced growth in height.\cite{Lallet1} As a result,
the interface area decreases and the energy cost required for the island to
create interface (3) decreases with regard to the whole internal energy of
the island. This allows one to understand why, on rough substrates,
$(111)_{\rm YSZ}$ rounded islands are observed only at the location of
defects.


\section{Conclusion}
\label{conclusion}

We have demonstrated using an {\em ab initio} approach that the $(111)_{\rm
YSZ} || (0001)_{\alpha-{\rm{}Al}_2{\rm{}O}_3}$ interface is thermodynamically
more stable than $(100)_{\rm YSZ} || (0001)_{\alpha-{\rm{}Al}_2{\rm{}O}_3}$ .
This result allows us to understand and coherently describe the islanding
process during the thermal treatment of YSZ on
$\alpha-{\rm{}Al}_2{\rm{}O}_3$, either perfect or with defects. On a perfect
substrate, the formation of the $(100)_{\rm YSZ} ||
(0001)_{\alpha-{\rm{}Al}_2{\rm{}O}_3}$ interface is energetically favored
over $(111)_{\rm YSZ} || (0001)_{\alpha-{\rm{}Al}_2{\rm{}O}_3}$, opening up
the way to the formation of $(100)_{\rm YSZ}$ islands on the perfect
substrate. On an imperfect surface, the formation of islands at the location
of defects leads to enhanced growth in height. As the interface area
decreases, the energy cost required to form interface (3) does too. As a
consequence, the $(111)_{\rm YSZ} || (0001)_{\alpha-{\rm{}Al}_2{\rm{}O}_3}$
interface can form at the location of defects, which explains that both
$(100)_{\rm YSZ}$ and $(111)_{\rm YSZ}$ islands are observed in this case.


\section*{Acknowledgements}

This work has been supported by grants from the Natural Sciences and
Engineering Research Council of Canada (NSERC) and the \textit{Fonds
Qu\'{e}b\'{e}cois de la Recherche sur la Nature et les Technologies} (FQRNT).
We are indebted to the \textit{R\'{e}seau Qu\'{e}b\'{e}cois de Calcul de
Haute Performance} (RQCHP) for generous allocations of computer resources. We
are grateful to Michel C\^ot\'e, Simon Pesant, Guillaume Dumont for help with the
{\em Abinit} code, and Michel B\'eland for advice on code optimisation.


\end{document}